\documentclass{elsart}
\usepackage{graphicx}

\begin{document}
\begin{frontmatter}
\title{Exciton scattering in light-harvesting systems of  purple bacteria}
 \author[HK]{Pavel He\v{r}man\thanksref{ph}},
\author[Chemnitz]{ Ulrich  Kleinekath\"ofer},
\author[Prague]{Ivan Barv\'{\i}k},
\author[Chemnitz]{Michael Schreiber} 
\thanks[ph]{Fax: +42 049 5061186, e-mail:
 pavel.herman@uhk.cz}

\address[HK]{Department of Physics, University of Hradec Kr\'{a}lov\'{e},
 V. Nejedl\'{e}ho 573, CZ-50003 Hradec Kr\'{a}lov\'{e}, Czech Republic}  
\address[Chemnitz]{Institut f\"ur Physik, Technische Universit\"at, 
D-09107 Chemnitz, Germany}
\address[Prague]{Institute of Physics of Charles University, Faculty of Mathematics and Physics,
 CZ-12116 Prague, Czech Republic }

\journal{Journal of Luminescence}
\date{\today}

\begin{abstract}
  Using the reduced density matrix formalism the exciton scattering in
  light-harvesting systems of purple bacteria is calculated.  The static
  disorder (fluctuations of the site energies) as well as the dynamic
  disorder (dissipation) is taken into account in this work.  Circular
  aggregates with 18 pigments are studied to model the B850 ring of
  bacteriochlorophylls with LH2 complexes.  It can be shown that the
  influence of dissipation may not be neglected in the simulation of the
  time-dependent anisotropy of fluorescence. Also an elliptical deformation
  of the ring could be essential.
\end{abstract}


\end{frontmatter}

\newcommand{\be}{\begin{equation}}
\newcommand{\ee}{\end{equation}}
\newcommand{\bea}{\begin{eqnarray}}
\newcommand{\eea}{\end{eqnarray}}
\newcommand{\ba}{\begin{array}}
\newcommand{\ea}{\end{array}}
\newcommand{\ok}{\omega_k}
\newcommand{\skp}{\sin{\frac{k}2}}
\newcommand{\svisle}{|G_k^1-G_k^2|^2}
\newcommand{\bose}{[1+2n_B(\hbar \ok)]}
\newcommand{\podil}{\frac{\cosh(w)}{\sinh(w)}}
\newcommand{\odmocnina}{\sqrt{1-u^2}}
\newcommand{\imag}{ {\rm i } }
\newcommand{\ddt}{\frac{\text d}{{\rm d} t}}

\section{Motivation}

Highly efficient light collection and excitation transfer to the reaction
center initiates the energy conversion in photosynthesis.  This process
takes place in the so-called light-harvesting antenna network.
Particularly, the ring subunit of the peripheral light-harvesting antenna
(LH2) \cite{dermott} of purple bacteria has been extensively studied
\cite{grond}.  The very symmetric arrangement with short distances between
the pigments gave new impulses to the discussion about coherence in the
exciton transfer and the exciton delocalization in LH2. Both could be
reduced by dynamic and static disorders.

Time-dependent experiments made it possible to study the femtosecond
dynamics of the energy transfer and relaxation \cite{nag1}.  Kumble and
Hochstrasser \cite{hoch} have presented a time-domain analysis of the
effects of the static disorder upon the dynamics of optical excitations.
The interpretation of time-dependent experiments on the femtosecond time
scale requires a theory which incorporates static and dynamic disorder.
The aim of the present paper is to extend the investigation by Kumble and
Hochstrasser taking into account the simultaneous influence of static and
dynamic disorders after impulse excitation. In addition we calculate the
time-dependent anisotropy of fluorescence not only for the symmetric but
also for elliptically distorted rings.

\section{Model}

The Hamiltonian 
\begin{equation}
H=H_{\mathrm{ex}}^{0}+H_{\mathrm{ph}}+H_{\mathrm{ex-ph}}+H_{\mathrm{s}}+
H_{\mathrm{el}}
\end{equation}
describes  the transfer of a  single exciton with a transfer integral
$J$ along the 
ideal ring ($H_{\mathrm{ex}}^{0}$), the independent heat baths 
for each chromophore  ($H_{\mathrm{ph}}$),  the
 site--diagonal and linear interaction between the exciton and 
the bath ($H_{\mathrm{ex-ph}}$),
the static disorder ($H_{\mathrm{s}}$) with
 Gaussian distribution (standard deviation $\Delta$) and
elliptical distortion of the ideal ring ($H_{\mathrm{el}}$).

Diagonalization of the Hamiltonian $H_{\mathrm{ex}}^{0}$ of the ideal ring
leads to eigenstates $|k \rangle$ and eigenenergies $E_k = - 2 \, J \, \cos
k$.  For a symmetrical coplanar arrangement of site transition moments
$\vec \mu{} _{n}$ dipole-allowed transitions populate only the degenerate $k= \pm{}
1$ levels of the ideal ring.  If the ring is elliptically distorted the
eigenenergies of the Hamiltonian $H_{\mathrm{ex}}^{0}+H_{\mathrm{el}}$ are
no longer degenerate as is the case for the ideal ring.  With static
disorder of the site energies being present ($ \Delta \neq 0 $), the
stationary states, i.e.\ the eigenstates $| a\rangle$ of the Hamiltonian $
H_{\mathrm{ex}}^{0} + H_{\mathrm{s}}$, correspond to mixtures of $|k
\rangle $ and an excitation will prepare a superposition of the $|k
\rangle$ states.

The dipole strength $\vec\mu{} _{a}$ of state $|a\rangle$ of the ring with
static disorder and the dipole strength $\vec \mu{} _{\alpha}$ of state $
|\alpha\rangle$ of the ideal one read
 \begin{equation}
 \vec\mu{} _{a} = \sum_{n=1}^{N} c_{n}^{a} \vec \mu{} _{n},
\qquad  
\vec \mu{} _{\alpha} =\sum_{n=1}^{N} c_{n}^{\alpha} \vec \mu{} _{n}.
\end{equation}
  The coefficients $c_{n}^{\alpha}$ and $c_{n}^{l}$ are
the expansion coefficients of the eigenstates of the ideal and disordered
rings in site representation.

Kumble and Hochstrasser \cite{hoch} concluded, that in the case of pump
pulse excitation the dipole strength is simply redistributed among the
exciton levels due to disorder. So the amplitudes of site excitations and
the phase relationships in the initial state are necessarily identical to
that of an equal superposition of $ k=\pm{} 1$ excitons of the ideal ring.
Thus, generally, the excitation with a pump pulse of sufficiently wide
spectral bandwidth will always prepare the same initial state, irrespective
of the actual eigenstates of the real ring. The nature of this initial
state is entirely determined by the selection rules of the ring without
static disorder.  The initial condition for the density matrix by pulse
excitation with the polarization $\vec e_{x}$ is given by (Eq. (1a) in
\cite{nag2}):
\begin{equation} 
\rho_{\alpha\beta} (t=0; \vec e_{i})   = 
\frac{1}{A}( \vec e_{x}\cdot{} \vec \mu{} _{\alpha})
( \vec e_{x}\cdot{} \vec \mu{} _{\beta}),
\end{equation}
where $A=\sum_\alpha ( \vec e_{x}\cdot{} \vec \mu{} _{\alpha})
( \vec e_{x}\cdot{} \vec \mu{} _{\alpha})$.

\section{Anisotropy of fluorescence}

Kumble and Hochstrasser \cite{hoch} calculated the usual time-dependent
anisotropy of fluorescence 
 \begin{equation}
 \label{aniso}
 r(t) = \frac { \langle S_{xx}(t)\rangle - \langle S_{xy}(t)\rangle} 
{ \langle S_{xx}(t)\rangle + 2\langle S_{xy}(t)\rangle}
\end{equation}
where, e.g.,
\begin{equation}
\langle S_{xy}(t) \rangle =
\langle | \sum_{\alpha,l,n} (\vec{e}_x\cdot{} \vec\mu{} _{\alpha})
(\vec{e}_y\cdot{} \vec\mu{} _{l}) c_{n}^{\alpha *}c_{n}^{l} 
{\mathrm e}^{-i \omega_l t}|^{2} \rangle.
\end{equation}
The indices $\alpha$ and $l$ label the
eigenstates of the virtual and disordered ring, respectively.
 The brackets $ \langle \rangle$ denote the
ensemble average and the average over the direction of the laser pulses
with fixed relative directions $\vec e_x$ and $\vec e_y$.

To include the dynamic disorder which contributes to dephasing of 
the initial wave packet and promotes  thermalization of 
the dephased populations one has to work within the exciton density
matrix formalism \cite{her} instead of using only the exciton wave 
functions
\begin{equation}
S_{xy}(t)=\int P_{xy}(\omega,t)d\omega
\end{equation}
where
\begin{equation} \label{trf}
   P_{xy}(\omega,t)=
 A\sum\limits_{l} \sum\limits_{ l'}
 \rho_{ll'}(t) ( \vec e_{y}\cdot{} \vec \mu{} _{l'}) 
( \vec e_{y}\cdot{} \vec \mu{} _{l})
 [\delta(\omega-\omega_{l'0})+ \delta(\omega-\omega_{l0})]. 
\end{equation}

\section{Density matrix formalism for  exciton transfer and relaxation}

Provided that the exciton dynamics is not very fast, its coupling to the
bath  weak, and except for the initial time-interval
$t\stackrel{<}{\sim}t_d$ ($t_d$~=~dephasing time of the bath), the
adequate equation  for a factorized initial state 
is the  Redfield equation \cite{redfield1,redfield2}:
\begin{equation}
 i\frac{d}{dt}\rho(t)=\frac{1}{\hbar}[H,\rho(t)]+{\mathcal R}\rho(t),
\label{eqmo}
\end{equation}
The Redfield relaxation superoperator ${\mathcal R}$ describes 
the influence of the thermal bath on the dynamics of the exciton.

\v{C}\'{a}pek  applied several different ways of
obtaining of  convolutional and convolutionless dynamical equations for
the exciton density matrix in the  site basis \cite{cap}. 
After Markovian approximation they have  the  following form
\begin{equation}
 i\frac{d}{dt}\rho(t)=\frac{1}{\hbar}[H,\rho(t)]-\delta\Omega\rho(t).
\label{eqmo}
\end{equation}
We proved \cite{her} the equivalence of the Redfield theory without secular
approximation with \v{C}\'{a}pek's equations after Markovian
approximation.

\section{ Results}

In Kumble and Hochstrasser's modeling, the anisotropy of fluorescence of
the ring LH2 subunit decreases from $0.7$ to $0.3-0.35$ and subsequently
establishes a final value of $0.4$.  Kumble and Hochstrasser concluded that
one needs static disorder of strength $\Delta \approx 0.4 - 0.8 J$ to reach
a time decay below 100 fs .

Our results for the time dependence ($t=\tau \hbar/J$)
of the anisotropy of fluorescence Eq.\ (\ref{aniso}) in
the symmetrical ring are shown on Fig. 1 with static disorder
$\Delta/J=0.4$.  In addition we consider two strengths of dynamic disorder
$j_0= 0.2$, 0.4 entering the spectral density
$J(\omega)=\Theta(\omega)j_{0}\frac{\omega^{2}}{2 \omega_c^3} {\mathrm
  e}^{-\omega/\omega_c}$ with cut-off frequency $\omega_c=0.2 J$
\cite{may}.  Inclusion of dynamic disorder leads to faster decay of the
anisotropy of fluorescence during the initial stage. 
Smaller strength of the static
disorder $\Delta/J$ than predicted by Kumble and Hochstrasser would be
necessary to guarantee the decrease of the anisotropy of fluorescence 
during the first 100
fs.

It was concluded in Ref.~\cite{hoch} based on measurements by Chachisvilis
et al.~\cite{chach} that the time decay of the anisotropy of fluorescence 
during the first
dozens of fs is { \em temperature independent} in the case of LH2 subunits.
Our calculation for the symmetrical ring show that such result can be
obtained only for $\Delta >0.8$.  But because the time resolution of the
experiments in \cite{chach} was not too high this very restrictive
statement about the strength of the static disorder can only be made with
caution. We expect that some temperature dependence can be seen using
experiments with shorter laser pulses.

 Recent results obtained by single molecule
spectroscopy \cite{oijen} can only be interpreted \cite{ket}
admitting the presence of a $C_{2}$  distortion of the LH2 ring. It has,
up to now, not been concluded whether such $C_{2}$
distortion of the LH2 ring is present also in samples in vivo. 

We have made calculations which take into account the possible $C_{2}$
distortion of the ring using the model C by Matsushita et al. 
\cite{ket} of  elliptical distortion of the ring in which 
the transfer integral $J$ is cosine modulated. Local transition dipole 
moments lie tangentially to the ellipse.
Our results for the time dependence of the anisotropy of fluorescence 
(\ref{aniso}) 
in the elliptically distorted ring are  shown on Fig. 2
with the static disorder $\Delta/J=0.4$ and  the strength of the dynamic 
disorder $j_0=0.2$ for three values of the elliptical deformation
$V_2/J = 0.05$, 0.1, 0.2.

It is seen, that the inclusion of an elliptical deformation 
leads to  faster decay of the anisotropy of fluorescence and 
diminishes the influence of the dynamic disorder.
Our calculation for the elliptically distorted ring 
($V_2/J=0.2$ \cite{ket})  show that  
the { \em temperature independent} time decay of the anisotropy of 
fluorescence during
the first dozens of fs can  be obtained even  for $\Delta \approx 0.4$ .

\begin{ack}
  This work has been partially funded by the M\v{S}MT \v{C}R 
(Kontakt CZE0014), BMBF and DFG. While preparing
  this work, I.B.  and P.H. experienced the kind hospitality of the
  Chemnitz University of Technology and U.K. experienced the kind
  hospitality of the Charles University in Prague.
\end{ack}

\bibliographystyle{elsart-num}

\begin{thebibliography}{10}

\bibitem{dermott} G. McDermott, S.M. Prince, A.A. Freer, 
A.M. Hawthornthwaite-Lawiess, M.Z. Papiz, R.J. Cogdell, N.W. Issacs,
{\em Nature} {\bf 374} (1995) 517.

\bibitem{grond} V. S\"{u}ndstr\"{o}m, T. Pullerits and R. van Grondelle, 
{\em J. Phys. Chem.} {\bf B 103} (1999) 2327.

\bibitem{nag1} V. Nagarjan, R.G. Alden, J.C. Williams, W.W. Parson,
{\em Proc. Natl. Acad. Sci. USA} {\bf 93} (1996) 13774.

\bibitem{hoch} R. Kumble, R. Hochstrasser,
{\em J. Chem. Phys.} {\bf 109} (1998) 855.

\bibitem{nag2} V. Nagarjan, E.T. Johnson, J.C. Williams,
W.W.  Parson, {\em J. Phys. Chem.} {\bf B 103} (1999) 2297.

\bibitem{her} P. He\v{r}man, U. Kleinekath\"{o}fer, I. Barv\'{\i}k,
M. Schreiber, {\em Chem. Phys.} (submitted)

\bibitem{redfield1} A.G. Redfield,
{\em IBM J. Res. Dev.} {\bf 1} (1957) 19.

\bibitem{redfield2} A.G. Redfield,
{\em Adv. Magn. Reson.} {\bf 1} (1965) 1.

\bibitem{cap}V. \v{C}\'{a}pek,
{\em Z. Phys.} {\bf B 99} (1996) 261.

\bibitem{may} V.~May, O.~K\"{u}hn, {\em Charge and Energy Transfer 
Dynamics in Molecular Systems} (Wiley-VCH, Berlin, 2000).

\bibitem{chach} M. Chachisvilis, O. K\"{u}hn, T. Pullerits,  V.
S\"{u}ndstrom, {\em J. Phys. Chem.} {\bf B 101} (1997) 7275.

\bibitem{oijen} A. M. van Oijen, M. Ketelaars, J. K\"{o}hler, 
T. J. Aartsma, J. Schmidt,  {\em Science} {\bf 285} (1999) 400. 

\bibitem{ket} M. Matsushita, M. Ketelaars, A. M. van Oijen, J. K\"{o}hler, 
T. J. Aartsma, J. Schmidt,  {\em Biophys. J.} {\bf 80} (2001) 1604.




\end{thebibliography}



\newpage


\begin{figure}
\begin{center}
\includegraphics[width=7.5truecm]{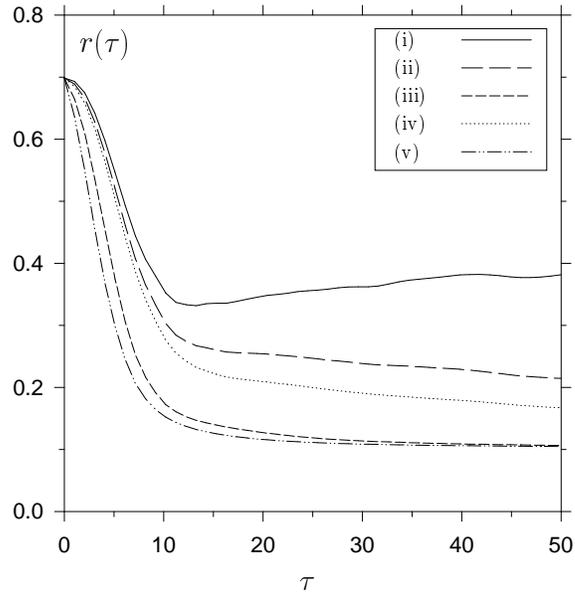}
\end{center}
\caption{
Time dependence of the anisotropy of fluorescence $r(\tau)$ for the symmetrical
ring with the static disorder $\Delta/J=0.4$. The influence of the dynamic 
disorder is displayed by curves for low (ii), (iv) and room temperature
(iii), (v) for $j_0=0.2$ (ii), (iii), and $j_0=0.4$ (iv), (v) compared to
$j_0=0.0$ (i).  
}
\end{figure}

\vspace{2cm}

\clearpage
\begin{figure}
\begin{center}
\includegraphics[width=7.5truecm]{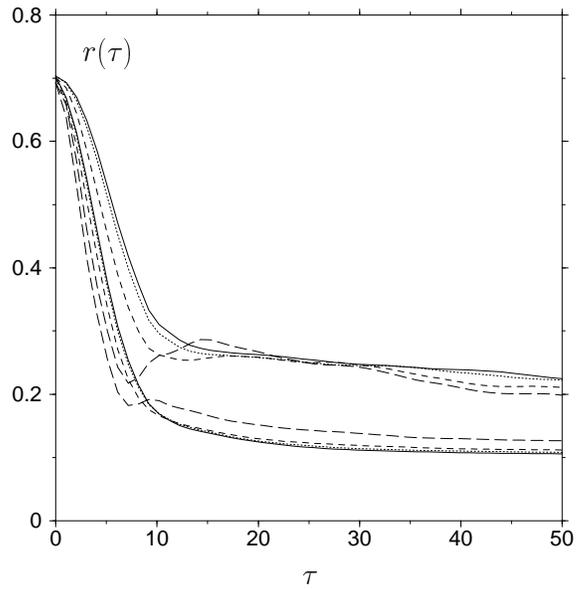}
\end{center}
\caption{
Time dependence of the anisotropy of fluorescence 
$r(\tau)$ for the elliptically 
distorted ring with the static disorder $\Delta/J=0.4$ and dynamic disorder 
$j_0=0.2$. The influence of the elliptical deformation displayed by dotted 
lines for $V_2/J=0.05$, dashed lines for $V_2/J=0.1$ and long dashed lines
for $V_2/J=0.2$
compared to $V_2/J=0.0$ (solid lines) for low and room temperature.
 }
\end{figure}

\end{document}